\documentclass[11pt,toc]{JHEP3}
\title{ Mass, Angular Momentum and Thermodynamics in Four-Dimensional
Kerr-AdS Black Holes}
\author{Rodrigo Olea \\
Departamento de F\'{\i}sica, Pontificia Universidad Cat\'{o}lica de Chile,\\
Casilla 306, Santiago 22, Chile.\\
E-mail: rolea@fis.puc.cl}

\abstract{In this paper, the connection between the
Lorentz-covariant counterterms that regularize the
four-dimensional AdS gravity action and topological invariants is
explored. It is shown that demanding the spacetime to have a
negative constant curvature in the asymptotic region permits the
explicit construction of such series of boundary terms.\newline
The orthonormal frame is adapted to appropriately describe the
boundary geometry and, as a result, the boundary term can be
expressed as a functional of the boundary metric, extrinsic
curvature and intrinsic curvature. This choice also allows to
write down the background-independent Noether charges associated
to asymptotic symmetries in standard tensorial formalism.\newline
The absence of the Gibbons-Hawking term is a consequence of an
action principle based on a boundary condition different than
Dirichlet on the metric. This argument makes plausible the idea of
regarding this approach as an alternative regularization scheme
for AdS gravity in all even dimensions, different than the
standard counterterms prescription.\newline As an illustration of
the finiteness of the charges and the Euclidean action in this
framework, the conserved quantities and black hole entropy for
four-dimensional Kerr-AdS are computed.}
\preprint{}

\begin{document}

\section{Introduction}

The construction of background-independent conserved quantities in AdS
gravity has attracted the attention of several authors in the recent
literature, especially in the context of AdS/CFT correspondence \cite%
{Maldacena,Witten}.

In the Ref.\cite{ACOTZ4} we consider the addition of the Euler term to the
four-dimensional AdS gravity action, that leads to a background-independent
charge definition by means of the Noether theorem for asymptotic symmetries.
For that case, the total action can be expressed in the language of
differential forms as%
\begin{equation}
I_{4}=\frac{l^{2}}{64\pi }\int\limits_{M}\hat{\epsilon}_{ABCD}\left( \hat{R}%
^{AB}+\frac{1}{l^{2}}e^{A}e^{B}\right) \left( \hat{R}^{CD}+\frac{1}{l^{2}}%
e^{C}e^{D}\right)  \label{IG4}
\end{equation}%
where $e^{A}=e_{\mu }^{A}dx^{\mu }$ is the vierbein (local orthonormal
frame) and $\hat{R}^{AB}=\frac{1}{2}\hat{R}_{\mu \nu }^{AB}dx^{\mu }\wedge
dx^{\nu }$ is the 2-form Lorentz curvature constructed up from the spin
connection $\omega ^{AB}=\omega _{\mu }^{AB}dx^{\mu }$ as $\hat{R}%
^{AB}=d\omega ^{AB}+\omega _{C}^{A}\omega ^{CB}$. The symbol $\hat{\epsilon}%
_{ABCD}$ is the totally antisymmetric Levi-Civita tensor and the Latin
indices run in the set $A=\{0,1,2,3\}$. The hatted curvatures stand for $4-$%
dimensional ones, the wedge product $\wedge $ between the differential forms
is understood and $l$ is the AdS radius.

The action (\ref{IG4}) does not require the addition of any boundary term to
cancel the divergences that appear in the evaluation of the Euclidean
continuation (see the discussion below). As the Euler term is quadratic in
the Riemann curvature in four dimensions, it coincides with the Gauss-Bonnet
term, so that the action (\ref{IG4}) in standard tensorial notation is%
\begin{equation}
I_{4}=-\frac{1}{16\pi }\int\limits_{M}d^{4}x\sqrt{-\hat{g}}\left( \hat{R}%
-2\Lambda +\frac{l^{2}}{4}(\hat{R}^{\mu \nu \sigma \rho }\hat{R}_{\mu \nu
\sigma \rho }-4\hat{R}^{\mu \nu }\hat{R}_{\mu \nu }+\hat{R}^{2})\right)
\label{IG4tensor}
\end{equation}%
with the cosmological constant $\Lambda =-3/l^{2}$. The coupling constant in
front of the Euler-Gauss-Bonnet term has been fixed demanding a novel
condition on the asymptotic curvature rather than the standard Dirichlet
condition on the metric. The crucial step is to assume that the spacetime
has a constant (negative) curvature at the boundary, that is,

\begin{equation}
\hat{R}_{\alpha \beta }^{\mu \nu }=-\frac{1}{l^{2}}\delta _{\lbrack \alpha
\beta ]}^{[\mu \nu ]}.  \label{Rdelta}
\end{equation}%
on $\partial M$. The above relation represents an asymptotic local condition
but does not impose any further restriction on the global topology of the
solution. For instance, Eq.(\ref{Rdelta}) is satisfied not only by
point-like configurations (black holes) but also by extended objects (black
strings, domain walls, etc.). The general character of this approach is
suitable to treat a wide family of solutions, from topological black holes
to Kerr-AdS and Taub-NUT/Bolt-AdS, with a single formula for the conserved
quantities.

The presence of the Euler term in the action does not modify the equations
of motion, but radically change the form of the conserved charges, canceling
the typical divergences in spacetimes with cosmological constant. When
evaluated for asymptotic Killing vectors, this formula renders a finite
value and recovers the correct results for a large variety of solutions \cite%
{ACOTZ4}. In particular, the charge formula obtained from this procedure
corrects the anomalous factor in the Komar's potential \cite{Katz, KBL} in a
background-independent framework.

For higher even dimensions ($D=2n$), it was shown in
Ref.\cite{ACOTZ2n} that the addition of the Euler term
(topological invariant constructed up with $n$ Lorentz curvatures)
always regularize the definition of conserved quantities in
asymptotically AdS spacetimes.

Unfortunately, in odd-dimensional AdS gravity there is a severe obstruction
to a similar construction because topological invariants of the Euler class
do not exist for $D=2n+1$. Therefore, one can only supplement the action
with boundary terms whose explicit form depends on the kind of boundary
condition under consideration.

In the context of the AdS/CFT correspondence, the regularization of the AdS
action using counterterms was carried out by Henningson and Skenderis \cite%
{Sk-He,Sk}. In this work, they presented a systematic procedure to
reconstruct asymptotically AdS spacetimes for a given data of the
boundary metric. The same algorithm obtains the explicit form of
the counterterms required by the finiteness of the stress tensor.
For dilatonic (super)gravity, the regularization of the action and
the holographic conformal anomaly were shown in \cite{N-O}. The
conserved quantities defined through the quasilocal stress tensor
\cite{Brown-York} for the regularized gravity action are
background-independent and have been computed for a number of
solutions. In particular, in five dimensions, the mass for
Schwarzschild-AdS black holes appears to be shifted in a constant
respect the Hamiltonian one. This constant is interpreted as the
Casimir energy of the corresponding boundary
CFT \cite{Ba-Kr}. A formula for the vacuum energy for any odd dimension $%
D=2n+1$ was proposed by Emparan, Johnson and Myers \cite%
{Emparan-Johnson-Myers} as an extrapolation from the results
computed up to seven dimensions. This is essentially due to the
technical difficulty to obtain the explicit form of the boundary
terms in high enough $D$, what makes the full series of
counterterms for any dimension still unknown.

In an alternative approach to deal with the above problem, a finite action
principle for odd-dimensional AdS gravity was achieved supplementing the
Einstein-Hilbert action by appropriate Lorentz-covariant boundary terms \cite%
{MOTZodd}. Apart from the condition on the asymptotic curvature (\ref{Rdelta}%
), we demand a \emph{holographic} condition on the extrinsic
curvature of the boundary in order to make the on-shell action
stationary. This boundary condition had been first introduced in
\cite{MOTZCS} to solve the problem of regularization of
Chern-Simons AdS gravity in higher odd dimensions. The procedure
carried out in \cite{MOTZodd} singles out the form of the boundary
term for a given dimension as a functional of the boundary
tensorial objects. The mass for static black holes calculated
through the Noether theorem also introduces a vacuum energy for
AdS spacetime, and explicitly verifies the expression conjectured in \cite%
{Emparan-Johnson-Myers} for all odd dimensions.

In this paper, we go back to the even-dimensional case and
explicitly construct the boundary terms that regularize the AdS
action following a strategy similar to the one implemented in
\cite{MOTZodd}. In the Refs. \cite{ACOTZ4,ACOTZ2n}, the procedure
that leads to the conserved quantities was carried out in first
order formalism, in terms of the tetrad and the spin connection.
Therefore, the final form of the charges has an explicit
dependence on both fields, what introduces an ambiguity in the
formula due to the arbitrariness in the choice of the orthonormal
frame. This fact also makes difficult the comparison to other
methods to compute conserved quantities in AdS gravity. Here, we
show that a suitable choice of the tetrad removes such ambiguity
and allow us to write down a tensorial expression for the charge
in terms of the boundary metric, the boundary Riemann tensor and
the extrinsic curvature.

\section{Lorentz-covariant Counterterms as Surface Terms in D=4}

In this section, we are interested in constructing an appropriate
Lorentz-covariant boundary term for four-dimensional AdS gravity, so that
the action has an extremum for arbitrary variations of the fields.

As we shall see below, this can be done integrating the surface term from
the variation of the action once proper boundary conditions are imposed.

Let us consider the standard Einstein-Hilbert action with negative
cosmological constant supplemented in a boundary term $B_{3}$. In the
language of the tetrad and the spin connection this is written as

\begin{equation}
I_{G}=\frac{1}{32\pi }\int\limits_{M}\hat{\epsilon}_{ABCD}(\hat{R}%
^{AB}e^{C}e^{D}+\frac{1}{2l^{2}}e^{A}e^{B}e^{C}e^{D})+\int\limits_{\partial
M}B_{3}.  \label{IGB3}
\end{equation}%
An arbitrary variation of fields $e^{A}$ and $\omega ^{AB}$ produces the
equations of motion for General Relativity and a surface term

\begin{equation}
\delta I_{G}=\int\limits_{M}\varepsilon _{A}\delta e^{A}+\varepsilon
_{AB}\delta \omega ^{AB}+d\Theta ,
\end{equation}%
where $\varepsilon _{A}$ is the Einstein equation,
\begin{equation}
\varepsilon _{A}=\hat{\epsilon}_{ABCD}\left( \hat{R}^{BC}+\frac{1}{l^{2}}%
e^{B}e^{C}\right) e^{D}
\end{equation}%
and the equation $\varepsilon _{AB}=0$ simply implies that the torsion must
vanish since the tetrad is invertible.

In order to obtain the field equations we need to perform an integration by
parts that gives the first contribution to the surface term $\Theta $
\begin{equation}
\Theta =\frac{1}{32\pi }\hat{\epsilon}_{ABCD}\delta \omega
^{AB}e^{C}e^{D}+\delta B_{3}  \label{ST4}
\end{equation}%
where the second one is coming from the variation of the boundary term in
Eq.(\ref{IGB3}).

\subsection{Adapted coordinates}

We consider a radial foliation of the spacetime, where the line element is
written in Gaussian (normal) coordinates

\begin{equation}
ds^{2}=\hat{g}_{\mu \nu }dx^{\mu }dx^{\nu
}=N^{2}(r)dr^{2}+h_{ij}(r,x)dx^{i}dx^{j}  \label{NormalC}
\end{equation}%
that are useful to describe the boundary geometry. The boundary is
located at a fixed value of $r=r_{0}$. The most natural choice of
the local orthonormal frame is adapting the tetrad to the
boundary, taking the block-diagonal decomposition

\begin{eqnarray}
e^{1} &=&Ndr  \label{e1} \\
e^{a} &=&e_{i}^{a}dx^{i}.  \label{ea}
\end{eqnarray}%
where we have separated the tangent space indices as $A=\{1,a\}$ and the
spacetime ones as $\mu =\{r,i\}$. The indices of the tangent space and the
spacetime are lowered and raised with $\eta _{AB}$ and $\hat{g}_{\mu \nu }$,
respectively. This preferred frame choice still preserves the rotational
Lorentz invariance on the boundary as a residual symmetry of the fields.

As torsion vanishes, the spin connection can be completely determined in
terms of the tetrad $\omega ^{AB}=\omega ^{AB}(e^{A})$,%
\begin{equation}
\omega _{\mu }^{AB}=-e^{B\nu }\nabla _{\mu }e_{\nu }^{A}
\end{equation}%
where $\nabla _{\mu }$ is the covariant derivative in the Christoffel
symbol. From the above relation, we can calculate the components $\omega
^{1a}$, that turn out to be related to the vielbein on the boundary by
\begin{equation}
\omega ^{1a}=-K_{i}^{j}e_{j}^{a}dx^{i}=-K^{a}  \label{normalomega}
\end{equation}%
where $K_{ij}$ is the extrinsic curvature that in normal coordinates (\ref%
{NormalC}) is given by

\begin{equation}
K_{ij}=-\frac{1}{2N}h_{ij}^{\prime }.  \label{Kdef}
\end{equation}%
Here, the prime stands for the derivative in the radial
coordinate. However, for the vielbein choice
(\ref{e1}),(\ref{ea}), the components of the Lorentz connection
$\omega ^{ab}$ are not expressed in terms of tensorial quantities
on the boundary. That is a problem if we are interested in
establishing a
connection to the usual tensorial formalism, that is, the boundary term $%
B_{3}$ to be expressible as a local function of the boundary metric $h_{ij}$%
, the extrinsic curvature $K_{ij}$ and the intrinsic curvature $%
R_{ij}^{kl}(h)$ (Riemann tensor of the boundary metric).

Then, the aim is constructing up Lorentz-covariant boundary terms
with the ingredients we have in the formalism of the spin
connection and vierbein. However, we cannot use directly the spin
connection, because it is not a vector for Lorentz
transformations. In order to restore the Lorentz covariance, we
define the second fundamental form (SFF) as the difference of two
spin connections at the boundary,
\begin{equation}
\theta ^{AB}=\omega ^{AB}-\bar{\omega}^{AB},  \label{SFFdef}
\end{equation}%
where $\omega ^{AB}$ is the dynamical field (the one that is varied to
obtain the corresponding field equation) and $\bar{\omega}^{AB}$ is a fixed
reference that lives only on the boundary. We are assuming that $\bar{\omega}%
^{AB}$ transforms in the same way as $\omega ^{AB}$ under the action of the
group $SO(3,1)$, but not under functional variations that act only on the
dynamical fields.

In the vicinity of the boundary ($r=r_{0}$), we can always write down a
product metric
\begin{equation}
ds^{2}=N^{2}(r)dr^{2}+\bar{h}_{ij}(x)dx^{i}dx^{j}
\end{equation}%
such that the matching condition is given by $\bar{h}%
_{ij}(x)=h_{ij}(r=r_{0},x)$ \cite{eguchi,Spivak,Choquet-Dewitt,Myers}. This
provides the definition of a cobordant geometry, whose connection $\bar{%
\omega}^{AB}$ on $\partial M$ satisfies
\begin{equation}
\begin{array}{cc}
\bar{\omega}^{1a}=0, & \bar{\omega}^{ab}=\omega ^{ab}.%
\end{array}
\label{omegabar}
\end{equation}%
This can be expressed as the fact that the spin connection coming from\ the
cobordant metric possesses only `tangential' components. Equivalently, the
SFF defined in Eq. (\ref{SFFdef}) has only `normal' components,
\begin{equation}
\begin{array}{cc}
\theta ^{1a}=-K_{i}^{a}dx^{i}, & \theta ^{ab}=0.%
\end{array}
\label{thetanormal}
\end{equation}%
We remark that choosing a cobordant geometry that is locally a product
metric on the boundary is only one possibility of a more general set of
matching conditions to recover the explicit form of the SFF (\ref%
{thetanormal}). We also stress that the connection $\bar{\omega}^{ab}$ needs
only to be specified on the boundary $\partial M$, where it agrees with the
dynamical field $\omega ^{ab}$. In this sense, the formalism presented here
conceptually differs from any background-dependent procedure, because the
latter requires the substraction of a vacuum configuration defined in the
entire manifold, not just on the boundary.

From the definition of the Lorentz curvature two-form, we obtain the
following decomposition for $\hat{R}^{AB}$ on the boundary
\begin{eqnarray}
\hat{R}^{1a} &=&D_{i}(\omega )\theta _{j}^{1a}dx^{i}\wedge dx^{j},
\label{Cod} \\
\hat{R}^{ab} &=&\left( \frac{1}{2}R_{ij}^{ab}+\theta _{i1}^{a}\theta
_{j}^{1b}\right) dx^{i}\wedge dx^{j}  \label{GaCod}
\end{eqnarray}%
where $R^{ab}=\frac{1}{2}R_{ij}^{ab}dx^{i}\wedge dx^{j}$ is the boundary
2-form curvature associated to $\omega ^{ab}$. Here, we have neglected the
components along $dr$ because the boundary $\partial M$ is defined for a
fixed $r=r_{0}$. Using the Eq.(\ref{thetanormal}) and the projection to the
basis space

\begin{equation}
\hat{R}_{\mu \nu }^{AB}=\hat{R}_{\mu \nu }^{\lambda \rho }e_{\lambda
}^{A}e_{\rho }^{B}  \label{projectione}
\end{equation}%
we see that the above relations are nothing but the standard Gauss-Coddazzi
decomposition of the Riemann tensor for a radial foliation (\ref{NormalC})%
\begin{eqnarray}
\hat{R}_{ij}^{rl} &=&-\frac{1}{N}\nabla _{\lbrack i}K_{j]}^{l},
\label{Coddazzi} \\
\hat{R}_{ij}^{kl} &=&R_{ij}^{kl}-K_{i}^{k}K_{j}^{l}+K_{i}^{l}K_{j}^{k}.
\label{GCtensor}
\end{eqnarray}%
Notice the change in the relative sign in Eq.(\ref{GCtensor}) with respect
to a timelike (ADM) foliation.

In sum, we have chosen an adapted coordinates frame to be able to express
the different components of the tetrad and the spin in terms of relevant
tensorial quantities on the boundary. We know that the boundary vierbein $%
e^{a}$ (sometimes also known as First Fundamental Form) is equivalent to the
boundary metric $h_{ij}$, the SFF corresponds to the extrinsic curvature $%
K_{ij}$ and the boundary curvature $R^{ab}$ is the boundary Riemann tensor $%
R_{ij}^{kl}$. Then, in practice, the introduction of a reference spin
connection is motivated by the need of eliminating the explicit dependence
of $B_{3}$ on the components $\omega ^{ab}$, because they are part of a
connection (Christoffel symbol) for the boundary metric.

\subsection{Integration of the Boundary Term in D=4}

In order to have a well-defined action principle (the action to be
stationary under arbitrary variations of the fields) we are going to
consider again the asymptotic condition (\ref{Rdelta}) to make the surface term (%
\ref{ST4}) vanish. Then, assuming that the spacetime has constant
negative curvature at $\partial M$
($\hat{R}^{AB}=-\frac{1}{l^{2}}e^{A}e^{B}$) and
developing the surface term along the different components, we have%
\begin{equation}
\Theta =\frac{l^{2}}{16\pi }\hat{\epsilon}_{1abc}\left( \delta \omega ^{1a}%
\hat{R}^{bc}+\delta \omega ^{ab}\hat{R}^{1c}\right) +\delta B_{3}
\label{ST4dec}
\end{equation}%
We introduced a Levi-Civita tensor for the boundary submanifold as $\hat{%
\epsilon}_{1abc}=-\epsilon _{abc}$, and with this notation the Eq.(\ref%
{ST4dec}) takes the form%
\begin{equation}
\Theta =\frac{l^{2}}{16\pi }\epsilon _{abc}\left[ \delta K^{a}\left(
R^{bc}-K^{b}K^{c}\right) +\delta \omega ^{ab}DK^{c}\right] +\delta B_{3}
\label{ST4y2}
\end{equation}%
where we have use the Gauss-Coddazzi relations (\ref{Cod},\ref{GaCod}). The
first term can be written as%
\begin{equation}
\epsilon _{abc}\delta K^{a}R^{bc}=\delta \left( \epsilon
_{abc}K^{a}R^{bc}\right) -\epsilon _{abc}K^{a}\delta R^{bc}
\end{equation}%
where the second contribution contains the variation $\delta R^{bc}=D(\delta
\omega ^{bc})$. Integrating by parts (and dropping the total derivative
because we are already at the boundary), we see that the term that comes out
is exactly the same (with opposite sign) as the third one in Eq.(\ref{ST4y2}%
). Therefore, we are able to integrate out the boundary term $B_{3}$,
demanding that the total surface term $\Theta $ vanishes

\begin{equation}
B_{3}=\frac{l^{2}}{16\pi }\epsilon _{abc}K^{a}\left( R^{bc}-\frac{1}{3}%
K^{b}\wedge K^{c}\right) .  \label{B3forms}
\end{equation}%
Finally, in standard tensorial notation we write down the above expression as

\begin{equation}
B_{3}=-\frac{l^{2}}{32\pi }d^{3}x\sqrt{-h}\delta _{\lbrack
i_{1}i_{2}i_{3}]}^{[j_{1}j_{2}j_{3}]}K_{j_{1}}^{i_{1}}\left(
R_{j_{2}j_{3}}^{i_{2}i_{3}}-\frac{2}{3}K_{j_{2}}^{i_{2}}K_{j_{3}}^{i_{3}}%
\right)  \label{B3tensors}
\end{equation}%
where $\delta _{\lbrack i_{1}i_{2}i_{3}]}^{[j_{1}j_{2}j_{3}]}$ is
the completely antisymmetrized product of Kronecker deltas in the
boundary indices.

It is already evident from the above expression that $B_{3}$ does not
contain any term proportional to $\sqrt{-h}K$ (Gibbons-Hawking term), where $%
K=K_{ij}h^{ij}$ is the trace of the extrinsic curvature \cite%
{Gibbons-Hawking}. This is not surprising since we are dealing with an
action principle different from the standard one based on a Dirichlet
boundary condition for the metric.

As we shall see below, this boundary term will provide a easy-to-use,
tensorial formula for the conserved charges in four dimensional AdS gravity
through the direct application of Noether theorem for asymptotic symmetries.

\section{Conserved Quantities}

\subsection{General Formula}

Let us consider an action that is the integral of a $D-$form Lagrangian
density in $D$ dimensions
\begin{equation}
L=\frac{1}{D!}L_{\mu _{1}...\mu _{D}}dx^{\mu _{1}}\wedge ...\wedge dx^{\mu
_{D}}.  \label{Ldform}
\end{equation}

An arbitrary variation $\bar{\delta}$ acting on the fields can be always
decomposed in a functional variation $\delta $ plus the variation due to an
infinitesimal change in the coordinates $x^{\prime \mu }=x^{\mu }+\eta ^{\mu
}$. For a $p$-form field $\varphi $, the latter variation is given by the
Lie derivative $\mathcal{L}_{\eta }\varphi $ along the vector $\eta ^{\mu }$%
, that can be written as $\mathcal{L}_{\eta }\varphi =(dI_{\eta }+I_{\eta
}d)\varphi $, where $d$ is the exterior derivative and $I_{\eta }$ is the
contraction operator \cite{Ichi}. The action of the functional variation $%
\delta $ on $L$ produces the equations of motion (Euler-Lagrange) plus a
surface term $\Theta (\varphi ,\delta \varphi )$ and, at the same time, the
Lie derivative contributes only with another surface term because $dL=0$.
Therefore, the Noether's theorem states that there exists a conserved
current associated to the invariance under diffeomorphisms of the Lagrangian
$L$, that is given by \cite{Choquet-Dewitt,Ramond}
\begin{equation}
\ast J=-\Theta (\varphi ,\delta \varphi )-I_{\eta }L.  \label{Jdiff}
\end{equation}%
For a diffeomorphism $\xi $ that is an isometry, the total variation $\bar{%
\delta}$ vanishes and then all the functional variations of the fields in $%
\Theta $ are replaced by the corresponding Lie derivative $\delta \varphi =-%
\mathcal{L}_{\xi }\varphi $ (see also \cite{Iyer-Wald} and, for a recent
discussion, \cite{H-I-M, HIM2}).

The conservation equation for the current $d\ast J=0$ expresses that $\ast J$
can always be written locally (by virtue of the Poincar\'{e}'s lemma) as a
total derivative. However, only when it can be written as an exact form $%
\ast J=dQ(\xi )$ globally, we can integrate the charge $Q(\xi )$ in a $%
(D-2)- $dimensional asymptotic surface (usually the boundary of the spatial
section, at constant time). This is exactly what we show below for the
Einstein-Hilbert action with the boundary term (\ref{B3forms}).

Plugging the boundary term $B_{3}$ into the expression for the total surface
term (\ref{ST4y2}), we obtain
\begin{equation}
\Theta =\frac{l^{2}}{16\pi }\epsilon _{abc}\delta K^{a}\left(
R^{bc}-K^{b}K^{c}+\frac{e^{b}e^{c}}{l^{2}}\right) ,  \label{Thetatotal}
\end{equation}%
that contains functional variations of the extrinsic curvature. In this
case, the current (\ref{Jdiff}) takes the explicit form%
\begin{equation}
\ast J=\frac{l^{2}}{16\pi }\epsilon _{abc}\left[ \mathcal{L}_{\xi
}K^{a}\left( R^{bc}-K^{b}K^{c}+\frac{e^{b}e^{c}}{l^{2}}\right) -I_{\xi
}DK^{a}\left( R^{bc}-K^{b}K^{c}+\frac{e^{b}e^{c}}{l^{2}}\right) \right]
\label{J+explicit}
\end{equation}%
where we have used the relation (\ref{Cod}) and dropped all the components
along $dr$. The Lie derivative on $K^{a}$ can be read off from the
corresponding components of the general expression for the spin connection%
\begin{equation}
\mathcal{L}_{\xi }\omega ^{AB}=DI_{\xi }\omega ^{AB}+I_{\xi }\hat{R}^{AB}
\label{LieOmega}
\end{equation}%
that, on the boundary $\partial M$, is simply written as%
\begin{equation}
\mathcal{L}_{\xi }K^{a}=DI_{\xi }K^{a}+I_{\xi }DK^{a}.  \label{LieK}
\end{equation}%
Finally, as the boundary torsion $T^{a}=De^{a}$ vanishes and the Bianchi
identity for the submanifold indices reads $D(R^{bc}-K^{b}K^{c})=0$, we are
able to write down the current as an exact form. We also assume the topology
of the manifold to be $R\times \Sigma $ (with $\Sigma $ as the spatial
section) and that the fields fall off rapidly enough to ensure the
convergence of the charge. Then, the conserved quantity associated to the
asymptotic symmetry $\xi $ is given by the integral at the boundary of the
spatial section

\begin{equation}
Q(\xi )=\frac{l^{2}}{16\pi }\int\limits_{\partial \Sigma }\epsilon
_{abc}I_{\xi }K^{a}\left( R^{bc}-K^{b}K^{c}+\frac{e^{b}e^{c}}{l^{2}}\right) .
\label{Qxi}
\end{equation}%
In standard tensorial notation, the charge (\ref{Qxi}) reads%
\begin{equation}
Q(\xi )=\frac{l^{2}}{32\pi }\int\limits_{\partial \Sigma }\sqrt{-h}\epsilon
_{i_{1}i_{2}i_{3}}\xi ^{k}K_{k}^{i_{1}}\left(
R_{mn}^{i_{2}i_{3}}-2K_{m}^{i_{2}}K_{n}^{i_{3}}+\frac{2}{l^{2}}\delta
_{m}^{i_{2}}\delta _{n}^{i_{3}}\right) dx^{m}\wedge dx^{n}
\label{Qxitensor4}
\end{equation}%
where now all indices are spacetime ones at the boundary and $dx^{m}\wedge
dx^{n}$ is the infinitesimal surface element of $\partial \Sigma $.

\subsection{Charges in four-dimensional Kerr-AdS}

The line element for the rotating solution in 4 dimensions can be written in
Boyer-Lindquist coordinates as \cite{carter}
\begin{equation}
ds^{2}=-\frac{\Delta }{\rho ^{2}}\left[ dt-\frac{a}{\Xi }\sin ^{2}\theta
d\phi \right] ^{2}+\frac{\rho ^{2}dr^{2}}{\Delta }+\frac{\rho ^{2}d\theta
^{2}}{\Delta _{\theta }}+\frac{\Delta _{\theta }\sin ^{2}\theta }{\rho ^{2}}%
\left[ adt-\frac{r^{2}+a^{2}}{\Xi }d\phi \right] ^{2},  \label{ds2kerr}
\end{equation}%
where the functions in the metric in terms of the spin parameter $a$ are%
\begin{eqnarray}
\Delta &\equiv &\left( r^{2}+a^{2}\right) \left( 1+\frac{r^{2}}{l^{2}}%
\right) -2mr,  \label{Deltar} \\
\Delta _{\theta } &\equiv &1-\frac{a^{2}}{l^{2}}\cos ^{2}\theta ,
\label{DeltaTheta} \\
\rho ^{2} &\equiv &r^{2}+a^{2}\cos ^{2}\theta ,  \label{rhodef} \\
\Xi &\equiv &1-\frac{a^{2}}{l^{2}}.  \label{bladef}
\end{eqnarray}%
Kerr-AdS black hole possesses an event horizon located at the radius $%
r=r_{+} $ such that it is the largest solution of the equation $\Delta
(r_{+})=0$.

The formula (\ref{Qxi}) for the vectors $\partial _{t}=\partial /\partial t$
and $\partial _{\phi }=\partial /\partial \phi $ for a rotating black hole,
evaluated on the sphere $S^{2}$ for $r\rightarrow \infty $, gives the results%
\begin{eqnarray}
Q\left( \partial _{t}\right) &=&\frac{m}{\Xi },  \label{E1} \\
Q\left( \partial _{\phi }\right) &=&\frac{ma}{\Xi ^{2}}  \label{J}
\end{eqnarray}%
where $Q\left( \partial _{\phi }\right) $ corresponds to the angular
momentum $J$. However, the first quantity $\tilde{E}=Q\left( \partial
_{t}\right) $ cannot be regarded as the energy for the Kerr-AdS black hole,
because the Killing field $\partial _{t}$ is rotating even at radial
infinity. In Boyer-Lindquist coordinates, the nonrotating timelike Killing
vector is the combination $\partial _{t}+\left( a/l^{2}\right) \partial
_{\phi }$, that substituted in the charge formula (\ref{Qxi}) gives the
\emph{physical }energy $E$%
\begin{equation}
E=Q\left( \partial _{t}+\frac{a}{l^{2}}\partial _{\phi }\right) =\frac{m}{%
\Xi ^{2}},  \label{Ephys}
\end{equation}%
in agreement with different methods in the literature \cite%
{H-TAdS,A-D,Caldarelli,G-P-P,D-K,B-C,note2}. The relevance of this result
has been emphasized by Gibbons, Perry and Pope in the context of the first
law of black hole thermodynamics \cite{G-P-P}, that is not satisfied by
other expressions for the Kerr-AdS energy previously found in the literature
\cite{HHP,Silva}.

\section{Black Hole Thermodynamics in Four Dimensions}

In this section, we use the boundary term (\ref{B3tensors}) to cancel the
divergences at radial infinity that appear in the explicit evaluation of the
bulk Euclidean action. For a given black hole solution, the Euclidean
continuation considers the horizon as shrunk to a point at the origin. The
requirement that the solution be smooth at the horizon fixes the period of
the Euclidean time $\beta $ (the inverse of the temperature $T$).

\subsection{Kerr-AdS}

We illustrate the finiteness of the Euclidean action with the addition of $%
B_{3}$ (\ref{IGB3}) for Kerr-AdS black hole as a nontrivial example. In this
case, the Euclidean period is given by the expression
\begin{equation}
\beta =T^{-1}=\frac{4\pi \left( r_{+}^{2}+a^{2}\right) }{r_{+}\left( 1+\frac{%
a^{2}}{l^{2}}+3\frac{r_{+}^{2}}{l^{2}}-\frac{a^{2}}{r_{+}^{2}}\right) }.
\label{beta}
\end{equation}%
The angular velocity of the black hole is%
\begin{equation}
\Omega =\frac{a\left( 1+\frac{r_{+}^{2}}{l^{2}}\right) }{r_{+}^{2}+a^{2}}
\label{Omega}
\end{equation}%
that is measured respect to a frame that is not rotating at infinity \cite%
{G-P-P}.

In the canonical ensemble, the Euclidean action $I_{E}$ is given by the free
energy, $I_{E}=\beta F$ that satisfies the thermodynamic relation

\begin{equation}
E-TS-\Omega J=TI_{E}  \label{thermo}
\end{equation}%
and defines the energy $E$ and the entropy $S$ of a black hole for a fixed
surface gravity (temperature) and angular velocity on the horizon.

The evaluation for Kerr-AdS metric of the Wick-rotated version of the action
(\ref{IGB3}) produces the finite value%
\begin{equation}
I_{E}=\frac{\pi \left( r_{+}^{2}+a^{2}\right) ^{2}\left( 1-\frac{r_{+}^{2}}{%
l^{2}}\right) }{\Xi l^{2}\left( \frac{3r_{+}^{4}}{l^{2}}+\left( 1+\frac{a^{2}%
}{l^{2}}\right) r_{+}^{2}-a^{2}\right) }  \label{EactB3}
\end{equation}%
in agreement with the standard result in the literature.  The divergences at
radial infinity in the bulk action are exactly canceled by the contributions
from the boundary term (\ref{B3tensors}) at $r=\infty $ (notice a sign
change because of the boundary orientation). At this point, we stress that,
once the temperature is fixed (in order to avoid the presence of a conical
singularity), this procedure requires neither the introduction of the
horizon as a new boundary nor ad-hoc boundary conditions on it, as claimed
in \cite{aros}.

Finally, with the expressions for the energy $E$ (\ref{Ephys}) and angular
momentum $J$ (\ref{J}) obtained in the previous section, the Eq. (\ref%
{thermo}) gives the standard result for the black hole entropy

\begin{equation}
S=\pi \frac{r_{+}^{2}+a^{2}}{\Xi }=\frac{1}{4}Area.  \label{entropy}
\end{equation}

\section{Discussion}

The standard counterterms approach considers boundary terms that
are local functional of the boundary metric $h_{ij}$ and Riemann
tensor $R_{ij}^{kl}$ and its covariant derivatives $\nabla
_{m}R_{ij}^{kl}$, and provides a systematic method to construct
them. However, in practice, for a given dimension the number of
possible counterterms increase drastically as we study more
complex solutions. Moreover, the extra terms needed for the
convergence of the stress tensor and the Euclidean action do not
seem to obey any particular pattern \cite{Das-Mann}. In that
spirit, one might naturally wonder if there is any other (more
compact) counterterms series that also regularize the AdS action.
We have shown in the previous sections that it is indeed possible
to construct such series of Lorentz-covariant counterterms. This
action principle might also provide some physical insight on how
to remove the ambiguities present in the standard counterterms
method in certain cases \cite{N-O2}.

Having the explicit form of the SFF (\ref{thetanormal}), we can always write
down the boundary term (\ref{B3forms}) in a fully Lorentz-covariant way

\begin{equation}
B_{3}=\frac{l^{2}}{32\pi }\hat{\epsilon}_{ABCD}\theta ^{AB}\left( R^{CD}+%
\frac{1}{3}\theta _{F}^{C}\theta ^{FD}\right) .  \label{B3cov}
\end{equation}%
The reader can be convinced that any linear combination of the above terms
and the expression $\hat{\epsilon}_{ABCD}\theta ^{AB}e^{C}e^{D}$ (the
fully-covariant version of the Gibbons-Hawking term) exhausts all possible
Lorentz-covariant boundary terms for 4-dimensional gravity constructed up
with the Levi-Civita as invariant tensor (and therefore, with the same
parity as the bulk terms in the Einstein-Hilbert action). Logically, the
term containing the tetrad does not appear in the final form of $B_{3}$,
because its variation would include $\delta e^{a}$, what would necessarily
lead us back to a Dirichlet condition for the boundary metric.

The fully-covariant expression for $B_{3}$ coincides with the boundary term
present in the Euler theorem \cite{eguchi}%
\begin{equation}
\int\limits_{M}\hat{\epsilon}_{ABCD}\hat{R}^{AB}\hat{R}^{CD}=32\pi ^{2}\chi
(M)+2\int\limits_{\partial M}\hat{\epsilon}_{ABCD}\theta ^{AB}\left( R^{CD}+%
\frac{1}{3}\theta _{F}^{C}\theta ^{FD}\right)  \label{EulerT}
\end{equation}%
where $\chi (M)$ stands for the Euler characteristic of the manifold $M$. As
$\chi (M)$ is a topological number (a constant), the above relation simply
means that --from the dynamical point of view-- a variation of the Euler
term in the l.h.s is equivalent to the variation of the boundary term $B_{3}$%
. This boundary term can also be regarded as a transgression form
for the Lorentz group, an extension of a Chern-Simons form to
include an additional field, such that the result is truly gauge
invariant \cite{eguchi,TF}.

In the previous section, we used the boundary term $B_{3}$ to regularize the
Euclidean bulk action. However, we can also consider the regularizing effect
of the Euler term in the bulk, evaluating the Euclidean continuation of the
action (\ref{IG4tensor}). In this case, the Euclidean action appears just
shifted in a constant respect to the expression (\ref{EactB3})%
\begin{equation}
I_{E}^{4}=I_{E}+\pi l^{2}  \label{Ishift}
\end{equation}%
that, as a consequence, produces the entropy $S^{\prime }$
\begin{equation}
S^{\prime }=\frac{1}{4}Area+\pi l^{2}.  \label{Sshifted}
\end{equation}%
This constant is irrelevant for the thermodynamic description of the system,
but has a clear geometrical meaning. In fact, plugging the
Euler-Gauss-Bonnet term from Eq.(\ref{EulerT}) into the action (\ref%
{IG4tensor}) we find that both approaches are equivalent up to an \emph{%
integration} constant given in terms of the Euler characteristic as $\frac{%
\pi l^{2}}{2}\chi (M)$. This feature already is present in the evaluation of
the entropy for topological Schwarzschild-AdS black holes

\begin{equation}
ds^{2}=-\Delta (r)^{2}dt^{2}+\frac{dr^{2}}{\Delta (r)^{2}}+r^{2}d\Sigma
_{\gamma }^{2}  \label{static}
\end{equation}%
with $\Delta ^{2}=\gamma -\frac{2G\mu }{r}+\frac{r^{2}}{l^{2}}$. These
solutions posses a transversal section $\Sigma _{\gamma }$ of constant
curvature $\gamma =+1,0,-1$ such that when we used the expression for the
regularized action (\ref{IG4tensor}), the entropy gets an additional
contribution $\pi l^{2}\gamma $.

For higher even-dimensional AdS gravity ($D=2n$), a well-defined action
principle was found in \cite{ACOTZ2n} supplementing the action in the Euler
term
\begin{equation}
\mathcal{E}_{2n}=\hat{\epsilon}_{A_{1}...A_{2n}}\hat{R}^{A_{1}A_{2}}...\hat{R%
}^{A_{2n-1}A_{2n}}  \label{Euler2ndef}
\end{equation}%
and fixing its weight factor demanding the same boundary condition on the
asymptotic curvature as in the four-dimensional case. The topological
invariant $\mathcal{E}_{2n}$ again cancels the divergences coming from the
bulk action, such that the regularized action can be written as%
\begin{equation}
I_{2n}=-\frac{1}{16\pi }\int\limits_{M}d^{2n}x\sqrt{-\hat{g}}\left( \hat{R}%
-2\Lambda +\alpha _{2n}\delta _{\lbrack \mu _{1}...\mu _{2n}]}^{[\nu
_{1}...\nu _{2n}]}\hat{R}_{\nu _{1}\nu _{2}}^{\mu _{1}\mu _{2}}...\hat{R}%
_{\nu _{2n-1}\nu _{2n}}^{\mu _{2n-1}\mu _{2n}}\right)  \label{I2nreg}
\end{equation}%
where the cosmological constant is $\Lambda =-\frac{(D-1)(D-2)}{2l^{2}}$ and
the coupling constant of the Euler term is
\begin{equation}
\alpha _{2n}=(-1)^{n}\frac{l^{2(n-1)}}{2^{n}n\left[ 2(n-1)\right] !}.
\label{couling2n}
\end{equation}%
If we are interested in constructing explicitly the boundary term for this
case, we have to consider the Einstein-Hilbert action in differential forms
language%
\begin{equation}
I_{G}=\frac{1}{16\pi (D-2)!}\int\limits_{M}\hat{\epsilon}_{A_{1}...A_{2n}}%
\left( \hat{R}^{A_{1}A_{2}}e^{A_{3}}...e^{A_{2n}}+\frac{D-2}{Dl^{2}}%
e^{A_{1}}...e^{A_{2n}}\right) +\int\limits_{\partial M}B_{2n-1}.
\label{IG2n}
\end{equation}%
Following a procedure identical as the one shown in Section 2.2, we are able
to integrate out $B_{2n-1}$ as%
\begin{eqnarray}
B_{2n-1} &=&(-1)^{n}\frac{l^{2(n-1)}}{8\pi (D-2)!}\int\limits_{0}^{1}dt%
\epsilon _{a_{1}...a_{2n-1}}K^{a_{1}}\left(
R^{a_{2}a_{3}}-t^{2}K^{a_{2}}\wedge K^{a_{3}}\right) ...  \nonumber \\
&&\qquad \qquad \qquad \qquad \qquad ...\left(
R^{a_{2n-2}a_{2n-1}}-t^{2}K^{a_{2n-2}}\wedge K^{a_{2n-1}}\right)
\label{B2n-1}
\end{eqnarray}%
where the factor $\left( R^{ab}-t^{2}K^{a}\wedge K^{b}\right) $ appears $%
(n-1)$ times and the integration over the continuous parameter $t\in \lbrack
0,1]$ gives the relative coefficients in the binomial expansion. This
boundary term can be also cast in tensorial form%
\begin{eqnarray}
B_{2n-1} &=&(-1)^{n+1}\frac{l^{2(n-1)}}{2^{n+2}\pi (D-2)!}%
\int\limits_{0}^{1}dtd^{2n-1}x\sqrt{-h}\delta _{\lbrack
i_{1}...i_{2n-1}]}^{[j_{1}...j_{2n-1}]}K_{j_{1}}^{i_{1}}\left(
R_{j_{2}j_{3}}^{i_{2}i_{3}}-2t^{2}K_{j_{2}}^{i_{2}}K_{j_{3}}^{i_{3}}\right)
...  \nonumber \\
&&\qquad \qquad \qquad \qquad \qquad \qquad ...\left(
R_{j_{2n-2}j_{2n-1}}^{i_{2n-2}i_{2n-1}}-2t^{2}K_{j_{2n-2}}^{i_{2n-2}}K_{j_{2n-1}}^{i_{2n-1}}\right)
\label{B2n-1tensor}
\end{eqnarray}%
with the use of the totally antisymmetric Kronecker delta. The
reader can notice by simple inspection of formula
(\ref{B2n-1tensor}) that the absence of the Gibbons-Hawking term
is not a particular property of the boundary term in four
dimensions (\ref{B3tensors}), but the general rule for all even
dimensions in this framework.

The Noether theorem provides the expression for the conserved quantities
associated to an asymptotic symmetry $\xi $, given by%
\newline
\begin{eqnarray}
Q(\xi ) &=&(-1)^{n+1}\frac{l^{2(n-1)}}{2^{n+2}\pi (D-2)!}\int\limits_{%
\partial \Sigma }\sqrt{-h}\epsilon _{i_{1}...i_{2n-1}}\xi
^{k}K_{k}^{i_{1}}\left( \hat{R}_{m_{1}m_{2}}^{i_{2}i_{3}}...\hat{R}%
_{m_{2n-3}m_{2n-2}}^{i_{2n-2}i_{2n-1}}+\right.   \nonumber \\
&&\qquad \qquad \qquad \qquad \qquad \qquad +\left( -1\right) ^{n}\frac{%
2^{n-1}}{l^{2(n-1)}}\left. \delta _{m_{1}}^{i_{2}}...\delta
_{m_{2n-2}}^{i_{2n-1}}\right) d\sigma ^{m_{1}...m_{2n-2}},  \label{Q2n}
\end{eqnarray}%
where $d\sigma ^{m_{1}...m_{2n-2}}=dx^{m_{1}}\wedge ...\wedge dx^{m_{2n-2}}$
is the infinitesimal surface element of $\partial \Sigma _{D-2}$.

As an example, we can compute the conserved quantities for a six-dimensional
Kerr-AdS black hole with a single rotation parameter \cite{HHP}%
\begin{eqnarray}
ds^{2} &=&-\frac{\Delta }{\rho ^{2}}\left[ dt-\frac{a}{\Xi }\sin ^{2}\theta
d\phi \right] ^{2}+\frac{\rho ^{2}dr^{2}}{\Delta }+\frac{\rho ^{2}d\theta
^{2}}{\Delta _{\theta }}+  \nonumber \\
&&+\frac{\Delta _{\theta }\sin ^{2}\theta }{\rho ^{2}}\left[ adt-\frac{%
r^{2}+a^{2}}{\Xi }d\phi \right] ^{2}+r^{2}\cos ^{2}\theta d\Omega _{2}^{2},
\label{KerrAdS6}
\end{eqnarray}%
where%
\begin{equation}
\Delta \equiv \left( r^{2}+a^{2}\right) \left( 1+\frac{r^{2}}{l^{2}}\right) -%
\frac{2m}{r},
\end{equation}%
and $d\Omega _{2}^{2}$ is the line element on unit $S^{2}$. The other
functions in the metric remain the same as in the four-dimensional case.

Evaluating Eq.(\ref{Q2n}) for the Killing fields $\partial _{t}$ and $%
\partial _{\phi }$in this metric, we have%
\begin{eqnarray}
Q\left( \partial _{t}\right)  &=&\frac{4\pi }{3}\frac{m}{\Xi },
\label{E1in6} \\
Q\left( \partial _{\phi }\right)  &=&\frac{2\pi }{3}\frac{ma}{\Xi ^{2}}
\label{J6}
\end{eqnarray}%
where $Q\left( \partial _{\phi }\right) $ corresponds to the
angular momentum $J$. As we had already pointed out in the
four-dimensional case, the timelike Killing vector that is not
rotating at infinity is  $\partial
_{t}+\left( a/l^{2}\right) \partial _{\phi }$, and produces the physical %
energy $E$%
\begin{equation}
E=Q\left( \partial _{t}+\frac{a}{l^{2}}\partial _{\phi }\right) =\frac{2\pi
}{3}\frac{m}{\Xi }\left( 1+\frac{1}{\Xi }\right) ,  \label{Ephys6}
\end{equation}%
in agreement with the first law of black hole thermodynamics \cite{G-P-P}.

The regularized Euclidean action (\ref{IG2n}) in six dimensions
for the
Kerr-AdS solution (\ref{KerrAdS6}) gives%
\begin{equation}
I_{6}=\beta \frac{\pi }{3\Xi }\left(
M-\frac{r_{+}^{2}}{l^{2}}\left( r_{+}^{2}+a^{2}\right) \right)
\label{I6}
\end{equation}%
that, using the energy (\ref{Ephys6}), the angular momentum
(\ref{J6})
and the angular velocity (\ref{Omega}), produces the correct value for the entropy%
\begin{equation}
S=\frac{2\pi ^{2}}{3\Xi }r_{+}^{2}\left( r_{+}^{2}+a^{2}\right)
\label{S6}
\end{equation}%
that has also been computed by several authors using different methods \cite%
{HHP,Das-Mann,Aw-Jo,G-P-P,D-K,B-C}.

Finally, in $D=2n$ we can also write down the fully
Lorentz-covariant version of the boundary term (\ref{B2n-1}) and
discover the connection with the Euler term (\ref{Euler2ndef})
through the Euler theorem in higher even dimensions. As a
consequence, again both procedures to compute the Euclidean action
(with the boundary term $B_{2n-1}$ or with $\mathcal{E}_{2n}$ as a
bulk term) are simply related by a topological number.

\section{Conclusions}

In this paper, we have explicitly constructed the Lorentz-covariant
counterterms that regularize the action for AdS gravity in four dimensions.
Our starting point was a well-defined action principle consistent with a
boundary condition for the asymptotic curvature. Certain choice of the
orthonormal frame allows us to write down the boundary term and the Noether
charges associated to asymptotic symmetries in standard tensorial formalism.

We have also explored the connection between these regularizing counterterms
and topological invariants. As the Euler term is dynamically equivalent to a
boundary term (by virtue of the Euler theorem), the divergences in the
Euclidean action can be equally canceled by the bulk or the surface term.

It is remarkable how a single boundary condition achieves a finite action
principle: the action is stationary under arbitrary variations of the fields
and the conserved charges and the Euclidean action are finite. On the
contrary to an action principle that relies on a Dirichlet condition for the
metric, where we can add any boundary term that is a functional of the
boundary metric, here we can only incorporate boundary terms whose
variations are compatible with the asymptotic condition (\ref{Rdelta}). This
simply means that even though we can always add surface terms to the action,
in general, the addition of an arbitrary boundary term will spoil the
boundary condition. This argument seems to explain why, in the end, we have
a quite restrictive action principle in spite of a general assumption for
the boundary condition.

More technically, the fact that the boundary term $B_{3}$ is constructed
using totally antisymmetric $3-$forms seems to rule out many of the terms
present in the standard counterterms series. For instance, the present
formalism cannot include any term that contains covariant derivatives of the
intrinsic curvature $\nabla _{m}R_{ij}^{kl}$ because they would be
automatically eliminated by the Bianchi identity.

In this paper, we have just given one explicit example in higher even
dimensions. However, it is expected that the same arguments about the
convergence of the Euclidean action and the conserved quantities hold for
any dimension $D=2n$. We hope to report this elsewhere.

Although the boundary term $B_{3}$ substantially differs in its form from
the standard counterterms in four dimensions \cite{Das-Mann}, it is clear
that both approaches cancel the same divergent powers in $r$ arising from
the bulk action. A comparison might be performed for a particular
coordinates choice, for instance, using the Fefferman-Graham expansion for
the metric \cite{F-G}, suitable to describe the conformal structure of an
asymptotically AdS spacetime%
\begin{equation}
ds^{2}=\frac{l^{2}}{4\rho ^{2}}d\rho ^{2}+\frac{g_{ij}\left( \rho ,x\right)
}{\rho }dx^{i}dx^{j}  \label{FG}
\end{equation}%
where the boundary is located at $\rho =0$ and
\begin{equation}
g_{ij}\left( \rho ,x\right)
=\stackrel{\scriptscriptstyle{(0)}}{g}_{ij}\left( x\right) +\rho
\stackrel{\scriptscriptstyle{(1)}}{g}_{ij}\left( x\right) +\rho
^{2}\stackrel{\scriptscriptstyle{(2)}}{g}_{ij}\left( x\right) +...
\label{expansiong}
\end{equation}%
and $\stackrel{\scriptscriptstyle{(0)}}{g}_{ij}$ is a given
boundary data for the metric. Indeed, a simple computation tells
us that the expansion of the
relevant components of the Riemann tensor (\ref{GCtensor}) reads%
\begin{equation}
\hat{R}_{ij}^{kl}=-\frac{1}{l^{2}}\delta _{\lbrack
ij]}^{[kl]}+\rho \left(
\stackrel{\scriptscriptstyle{(0)}}{R_{ij}^{kl}}+\frac{1}{l^{2}}\left(
\delta
_{\lbrack i}^{k}\stackrel{\scriptscriptstyle{(1)}}{g}_{j]}^{l}+\stackrel{%
\scriptscriptstyle{(1)}}{g}_{[i}^{k}\delta _{j]}^{l}\right) \right) +...
\label{Rsubdelta}
\end{equation}%
where the extra terms are increasing powers of $\rho $ and the indices of
the tensorial coefficients are raised and lowered with $\stackrel{%
\scriptscriptstyle{(0)}}{g}_{ij}$. This reflects how the boundary condition (%
\ref{Rdelta}), required to attain a finite action principle, is
automatically satisfied in the coordinates frame (\ref{FG}).

\acknowledgments

I wish to thank M. Ba\~{n}ados, G. Barnich, G. Giribet, G. Kofinas, O. Mi%
\v{s}kovi\'{c}, S. Theisen and J. Zanelli for helpful discussions. I am also
grateful to the organizers of Spring School on Superstring Theory for
hospitality at Abdus Salam ICTP, Trieste and to Prof. Emparan for
hospitality at Universitat de Barcelona. This work was funded by the grant
3030029 from FONDECYT.


\begin{thebibliography}{99}
\bibitem{Ichi} The action of the contraction operator $I_{\eta }$ over a $p$%
-form $\alpha _{p}=\frac{1}{p!}\alpha _{\mu _{1}}\ldots _{\mu _{p}}dx^{\mu
_{1}}\wedge \ldots \wedge dx^{\mu _{p}}$ is given by $I_{\eta }\alpha _{p}=%
\frac{1}{(p-1)!}\eta ^{\nu }\alpha _{\nu \mu _{1}}\ldots _{\mu
_{p-1}}dx^{\mu _{1}}\wedge \ldots \wedge dx^{\mu _{p-1}}$.
\bibitem{note2} Notice that this result differs from the energy in \cite%
{ACOTZ4}, obtained for a different choice of the Kerr-AdS metric.
\end{thebibliography}
\end{document}